\documentclass{article}
\usepackage{spconf,amsmath,graphicx}

\usepackage{etex}
\usepackage{mathtools}
\usepackage{amssymb}
\usepackage{amsmath}
\usepackage{algorithm,algorithmicx,algpseudocode}
\usepackage{pictex,graphicx}
\usepackage{cite}
\usepackage{array}
\usepackage[tight,footnotesize]{subfigure}

\title{Optimization of zero-delay mappings for distributed coding by deterministic annealing} 

\begin{document}

\name{Mustafa S. Mehmetoglu, Emrah Akyol, Kenneth Rose}
\address{Department of Electrical and Computer Engineering\\ University of California, Santa Barbara, CA, 93106\\Email:\{mehmetoglu, eakyol, rose\}@ece.ucsb.edu}

\maketitle

\begin{abstract}
This paper studies the optimization of zero-delay analog mappings in a network setting that involves distributed coding. The cost surface is known to be non-convex, and known greedy methods tend to get trapped in poor locally optimal solutions that depend heavily on initialization. We derive an optimization algorithm based on the principles of ``deterministic annealing", a powerful global optimization framework that has been successfully employed in several disciplines, including, in our recent work, to a simple zero-delay analog communications problem. We demonstrate strict superiority over the descent based methods, as well as present example mappings whose properties lend insights on the workings of the solution and relations with digital distributed coding.
\end{abstract}


\section{Introduction}
\label{sec:intro}

It is well known that in the case of a memoryless Gaussian source and an additive white Gaussian noise channel, under the mean squared error distortion, the asymptotic information theoretic bound is achievable by a zero-delay scheme \cite{goblick}. Although this property does not apply to general sources and channels \cite{tocode}, the simple structure of a coding scheme without long delays has made joint source-channel coding an attractive problem of practical importance. 

Zero-delay coding problems have been studied extensively in the literature (see eg. \cite{hekland, karlsson2010optimized, hu2011analog, chen2011zero}), but optimal coding schemes for zero delay distributed coding problems are not known in general. The non-convex cost surface renders greedy descent methods  \cite{emrah_dcc10} inefficient. In prior work \cite{mehmetoglu_itw} we proposed a method based on deterministic annealing \cite{da} to optimize zero delay codes for a point-to-point communication setting where the decoder has access to additional side information. The method presented here extends the approach to distributed settings, where optimization of multiple encoders poses significant additional challenges. The hidden interaction between separate encoders leads to interesting coding schemes, which may be interpreted as some or all encoders acting as side information for the others. 

In Section \ref{prelim}, we state the problem and briefly summarize the descent-based algorithm. The proposed approach is described in Section \ref{proposed}. Numerical results and examples are presented in Section \ref{experiments}, and conclusion in Section \ref{conclusions}.

\section {Problem Definition and the Greedy Approach}
\label{prelim}
\subsection{Problem Definition}

\begin{figure} 
  \centering
  \centerline{\includegraphics[width=8.5cm]{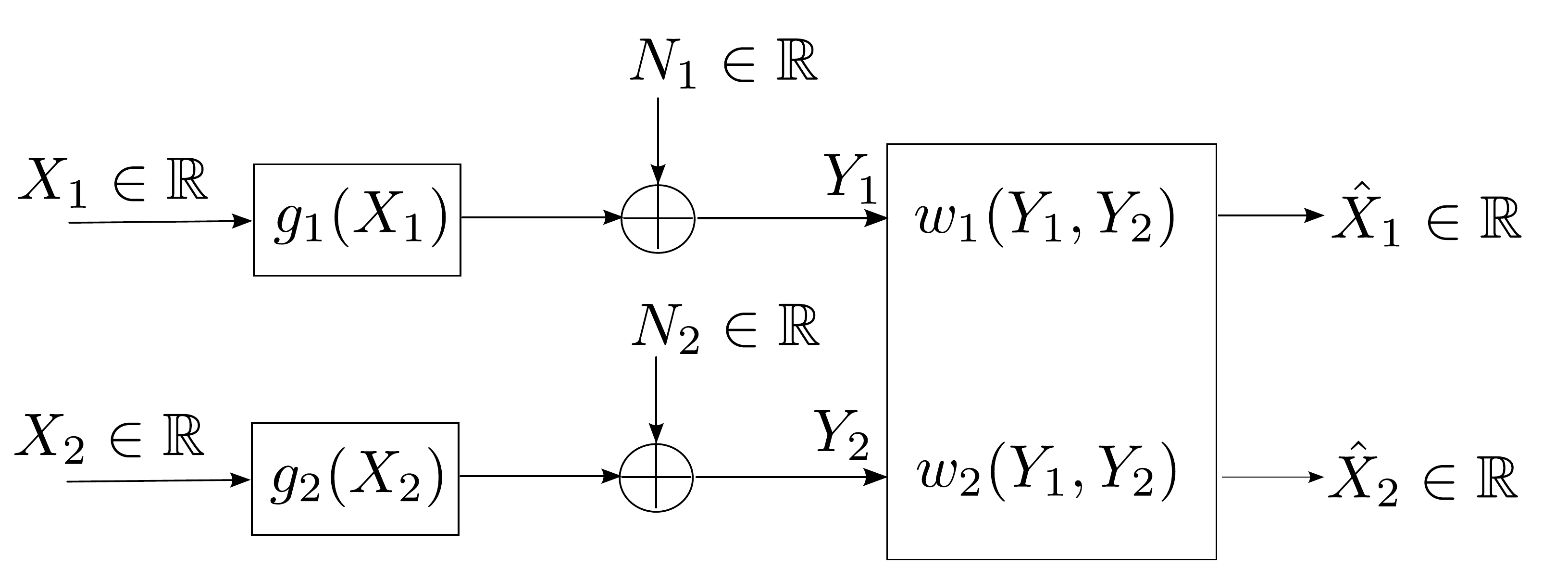}}
\caption{Problem Setting}
\label{distr}
\end{figure}

Let $\mathbb E\{\cdot\}$, $\mathbb P\{\cdot\}$ denote the expectation and probability operators, let  $\mathbb R$ be the set of real numbers, and $H(\cdot)$  be the Shannon entropy. The probability density function of the random variable $X$ is $f_{X}(x)$, where upper case letters are used to denote random variables and lower case letters for their realizations. $\nabla$ and $\nabla_x$ denote the gradient and partial gradient with respect to $x$, respectively. Logarithms in this paper are natural logarithms. 

The problem setting is given in Figure \ref{distr}, where two scalar sources ${X_1} \in  \mathbb R$ and ${X_2} \in  \mathbb R$ are drawn from joint density $f_{X_1,X_2}( \cdot,\cdot)$ and mapped to channel input by the encoding functions ${ g_1, g_2 }:\mathbb R\rightarrow \mathbb R$. Both channels have additive noises ${N_1, N_2} \in \mathbb R$ not necessarily independent from the sources and distributed according to $f_{N_1,N_2}(\cdot,\cdot)$. The decoders ${w_1}$ and ${w_2}$ map the received channel outputs ${Y_1}=g_1(X_1)+N_1$ and ${Y_2}=g_2(X_2)+N_2$ to the estimates ${X_1}$ and ${X_2}$. The problem is to find optimal mapping functions $ g_1 (\cdot), g_2(\cdot),w_1(\cdot,\cdot), w_2(\cdot,\cdot)$ that minimize the mean squared error (MSE) distortion
\begin{equation}
 D=\mathbb E\{(X_1-\hat X_1)^2 + (X_2-\hat X_2)^2\},
 \label{distortion1}
 \end{equation}
subject to power constraints on the encoders:
  \begin{equation}
 P_i \triangleq \mathbb E\{ g_i^2(X_i) \} \leq P_i', \text{ for } i=1,2.
 \label{power_cons1}
 \end{equation}
We formulate the problem as minimizing the Lagrangian cost given by
\begin{equation}
J=D+\lambda_1 P_1 +\lambda_2 P_2
\label{cost lagrangian}
\end{equation}
where $\lambda_1$ and $\lambda_2$ are Lagrange multipliers. Note that setting $\lambda_1 = \lambda_2 = \lambda$ would yield \begin{equation}
J=D+\lambda(P_1+P_2)
\end{equation}
 which can be interpreted as the ``total power constraint'' variant of the problem, where the constraint is given by $P_1+P_2<P'$. Although we focus on scalar sources and noises, our method can be extended to vector spaces albeit with more involved expressions.
 
\subsection {Greedy Algorithm}
\label{prior}
The optimal decoders given the encoders are the MSE estimators and given in closed form as:
\begin{equation}
w_i(y_1, y_2)= \mathbb E \{ X_i| y_1, y_2\}, \text{ for } i=1,2.
\label{decoder eq}
\end{equation}
The necessary conditions for optimality of encoders (given the decoders) are derived by requiring the functional derivative of the cost (\ref{cost lagrangian}) to vanish:
\begin{align}
 \nabla_{g_1} J [g_1,g_2]= \nabla_{g_2} J [g_1,g_2]=0, \, \, \forall \, x_1,x_2
\end{align} 
We omit detailed expressions for brevity, see \cite{emrah_dcc10} for full description. A greedy descent algorithm based on iterative imposition of necessary conditions for optimality was proposed in \cite{emrah_dcc10}. Since descent-based algorithms of this type are highly susceptible to getting trapped in poor local minima and heavily dependent on initialization, performance was improved by employing noisy channel relaxation (NCR) \cite{gadkari1999robust, Knagenhjelm}. As demonstrated in this work, NCR results are nevertheless suboptimal and further improvements are achievable.

\section {Proposed Method}
\label{proposed}
In our method we optimize encoders within a class of structured mappings that are defined in a piecewise manner via a space partition and a local model for each partition cell. We use affine mappings for each local model, which results in a piecewise linear function that approximates the desired optimal mapping. Note here that the choice of affine model is for simplicity and other, richer, local models (such as higher order polynomial) are possible. 

DA introduces randomness into the optimization process by randomizing the partition, i.e., points are associated in probability to partition cells and hence to local models. The randomness is measured by the Shannon entropy and is constrained while minimizing the expected cost. The resulting Lagrangian functional is referred to as ``free energy'' and Lagrange multiplier $T$ that controls the entropy term is called ``temperature", to emphasize an insightful analogy to statistical physics. The optimization process is akin to annealing of a physical system, starting at high temperature, where the cost is convex, and gradually lowering it to zero while minimizing the free energy at each temperature.  

\subsection {Encoder Mappings}
We approximate the encoders as piecewise linear mappings, where the first encoder is defined as $g_1(x)=g_{1,k_1}(x)$ for $x\in\mathbb R_{1,k_1}$ and for $k_1\in \{1,...,K_{1,max}\}$. Here the regions  denoted as $\mathbb R_{1,k_1}$ define the space partition, and $g_{1,k_1}$ are the parametric local models given by $g_{1,k_1}(x)=a_{1,k_1}x+b_{1,k_1}$. One can define $g_{1,k_1}$ differently to obtain different structures. The second encoder is defined similarly as $g_2(x)=g_{2,k_2}(x)$ for $x\in\mathbb R_{2,k_2}$ and $k_2\in \{1,...,K_{2,max}\}$.

Note that in the preceding definition, every input point is associated with one local model and the association is defined by the partitioning regions, hence the encoder outputs are deterministic. We now randomize the encoders by defining the following probabilities:
\begin{equation}
p(k_i|x_i) \triangleq \mathbb P\{ x_i \in \mathbb R_{i,k_i}\} , \quad \forall k_i,x_i, \text{ for }i=1,2.
\end{equation}

We write the cost in (\ref{distortion1}) accounting for the random encoders as:
\begin{equation}
 D=\mathbb E \{D_{K_1,K_2}(X_1,X_2)\}
 \label{distortion_new1}
 \end{equation}
 where expectation is taken over $\{X_1,X_2,K_1,K_2\}$ and $D_{k_1,k_2}(x_1,x_2)$ is given by   
 \begin{align}
 D_{k_1,k_2}&(x_1,x_2)\! \nonumber\\ 
 &=\!\mathbb E\{ (x_1\!-\!w_1(g_{1,k_1}(x_1)\!+\!N_1, g_{2,k_2}(x_2)\!+\!N_2) )^2 \nonumber \\
 &+\!(x_2\!-\!w_2(g_{1,k_1}(x_1)\!+\!N_1, g_{2,k_2}(x_2)\!+\!N_2) )^2 \}
 \label{pieceD}
 \end{align}
  Power constraints for encoders are written as:
 \begin{equation}
 P_i =\sum\limits_{k_i} \int\limits _{\mathbb R} g_{i,k_i}^2(x_i)p_{X_i}(x_i)p(k_i|x_i) \mathrm{d} x_i, \text{ for }i=1,2.
 \label{power_new1}
 \end{equation}  
 
\subsection {Entropy Constraint}
Note that if we minimize an unconstrained $J$ with respect to the association probabilities, the solution will be deterministic such that every input point is associated with probability one to the model that contributes the least to the cost. However, to mitigate local minima, we minimize $J$ subject to a constraint on the joint entropy of the system. We  construct the Lagrangian
\begin{equation}
F=J-TH
\label{free energy}
\end{equation}
or  ``free energy" to be minimized, with $T$ (temperature) the Lagrange multiplier controlling the entropy constraint. Noting the (by construction) Markov chain $K_1 \rightarrow X_1 \rightarrow X_2\rightarrow K_2$, we can express the joint entropy as  
 \begin{equation}
 H(X_1,K_1,X_2,K_2) = H(X,Y)+H(K_1|X_1) + H(K_2|X_2)
 \end{equation}
 Since $H(X,Y)$ is a constant determined by the sources, we define $H \triangleq H(K_1|X_1)+H(K_2|X_2)$ where
 \begin{equation}
 H(K_i|X_i) = -\int\limits _{\mathbb R}p_{X_i} (x_i)  \sum\limits_{k_i} p(k_i|x_i)\log(p(k_i|x_i))\mathrm{d} x_i
 \end{equation}
 for $i=1,2$.
 
  \subsection{Minimization of $F$} 
We optimize the free energy (\ref{free energy}) of the system with respect to encoders (association probabilities and local models) and decoders. It is easy to verify that optimal association probabilities are given by Gibbs distribution:
 \begin{align}
 p(k_1|x_1) &= \frac{e^{-[\mathbb E \{D_{k_1,K_2}(x_1,X_2)\} + \lambda_1 g_{1,k_1}^2(x_1)]/T} }{ \sum\limits_{k_1}e^{-[\mathbb E \{D_{k_1,K_2}(x_1,X_2)\} + \lambda_1 g_{1,k_1}^2(x_1)]/T} } \quad \forall x\\
  p(k_2|x_2) &= \frac{e^{-[\mathbb E \{D_{K_1,k_2}(X_1,x_2)\} + \lambda_2 g_{2,k_2}^2(x_2)]/T} }{ \sum\limits_{k_2}e^{-[\mathbb E \{D_{K_1,k_2}(X_1,x_2)\} + \lambda_2 g_{2,k_2}^2(x_2)]/T} } \quad \forall x.
 \label{optimum prob}
 \end{align}
The local models can be optimized through gradient descent search,  and the optimal decoders are given by (\ref{decoder eq}). Explicit expressions are omitted for brevity.

 \subsection{Algorithm}
The annealing process starts at a high temperature where the free energy (\ref{free energy}) is minimized through maximizing the entropy, which is achieved by uniform distribution. This means that all points are equally assigned to all models, which are therefore identical, and for each encoder we effectively have a single model. Note that, in agreement with this observation, as $T \rightarrow \infty$ in (\ref{optimum prob}) the associations probabilities become uniform. 

As we lower $T$, a temperature is reached where the present solution is no longer a minimum but a saddle point. A bifurcation occurs such that the local models are divided into two or more groups, the entropy is traded for reduction in the cost ($J$) and a lower free energy is obtained. Such bifurcations are referred to as ``phase transitions" and the particular temperature that they occur are called the ``critical temperatures". As $T \rightarrow 0$, minimizing the free energy is equivalent to minimizing $J$, which is achieved through deterministic encoders. At this point, the algorithm is equivalent to the greedy method described in Section \ref{prior}.

\begin{figure}
\includegraphics[width=8.5cm]{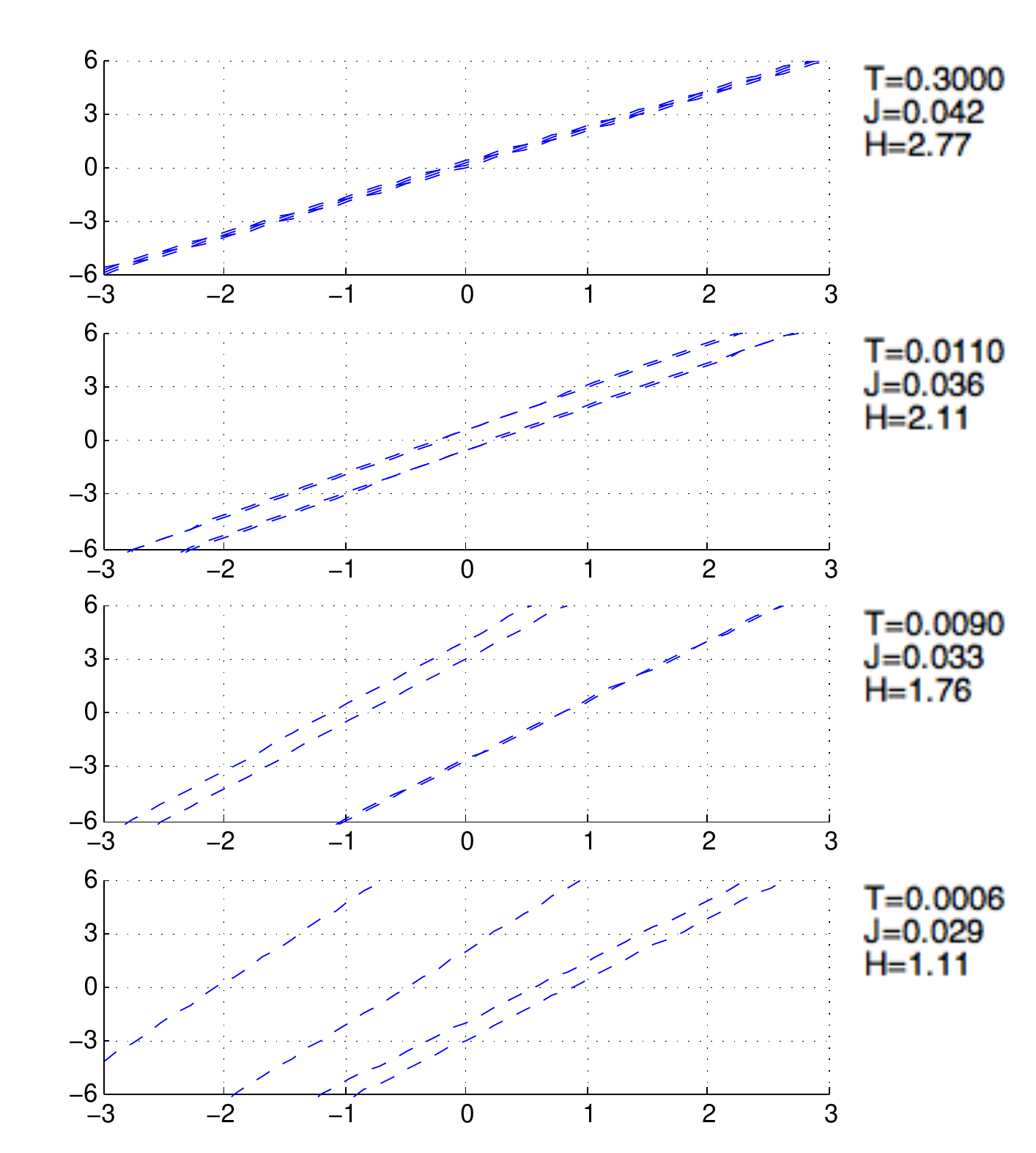}
\caption{Illustration of the method. The evolution of the first encoder is shown. }
\label{evolve}
\end{figure}

An illustration of the method is given in Figure \ref{evolve} where we show how one of the encoders evolves during annealing. Initially the 4 local models are coincident. Entropy is maximum at this temperature. Around $T=0.011$, the first critical temperature, the local models start dividing into two subgroups. At $T=0.009$, the second critical temperature (for this encoder), we see the first subgroup splits into two subgroups as well. Another phase transition is observed around $T=0.0006$. Further phase transitions can be obtained by creating duplicates of the local models. Note that we reduce the temperature in a geometric fashion, in Figure \ref{evolve} only the critical temperatures are shown. Moreover, in order to trigger a phase transition, the local models are perturbed slightly at every temperature. They split when we reach a new critical temperature and join back at others.

\section {Experimental Results}
\label{experiments}
In our experiments we used jointly Gaussian sources with unit variance and a correlation coefficient $\rho=0.995$. Noises are independent Gaussians with variance $\sigma^2=0.1$. 

\begin{figure}[htb]
\begin{minipage}[b]{1.0\linewidth}
  \centering
  \centerline{\includegraphics[width=8.5cm]{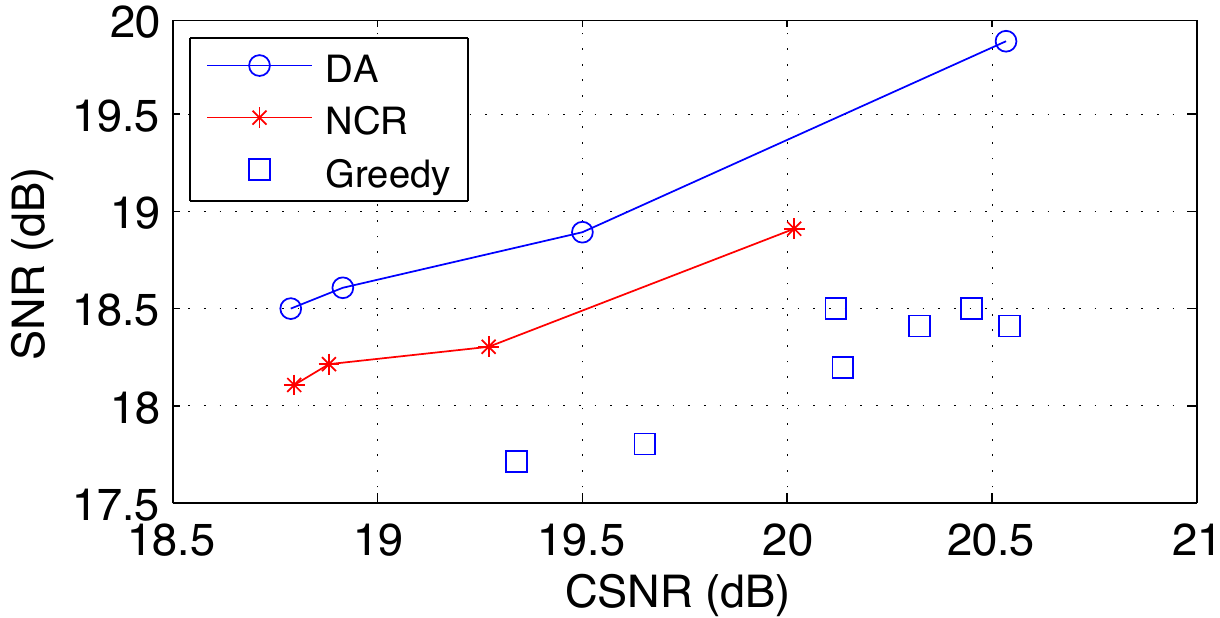}}
  \centerline{(a)}\medskip
\end{minipage}
\begin{minipage}[b]{1.0\linewidth}
  \centering
  \centerline{\includegraphics[width=8.5cm]{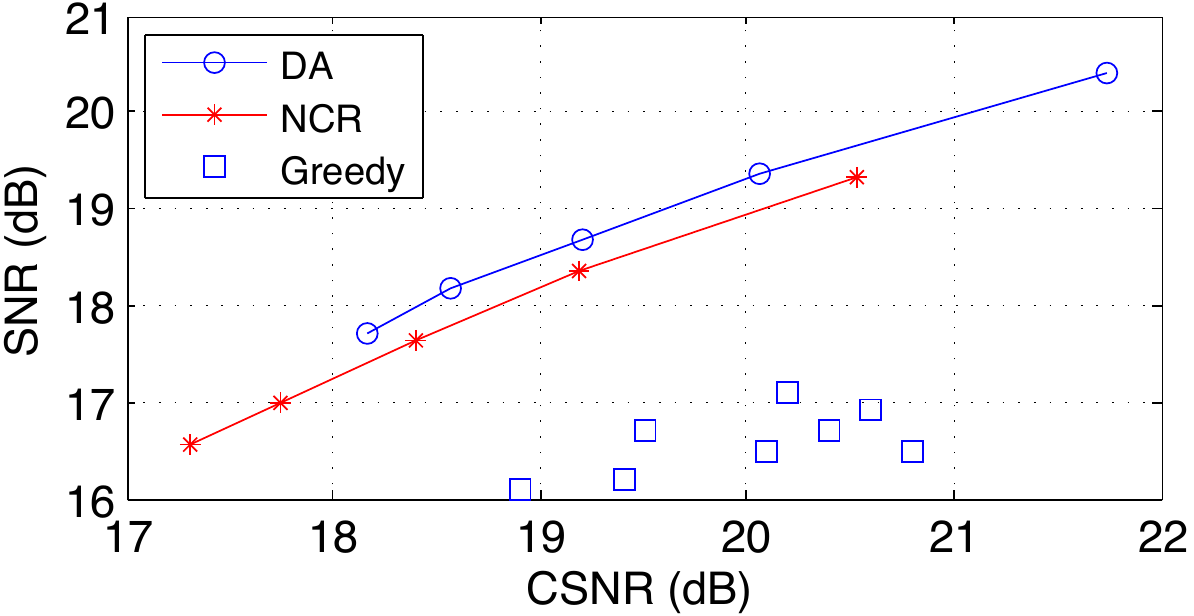}}
  \centerline{(b)}\medskip
\end{minipage}
\caption{Performance comparison plots. (a) Individual power constraints. (b) Total power constraint.}
\label{compp}
\end{figure}

Performance comparisons for individual power allocation case (different $\lambda$) and for total power allocation case (same $\lambda$) are provided in Figure \ref{compp} where we define $\text{SNR}=10\log_{10}(1/D)$ and $\text{CSNR}=10\log_{10}((P_1+P_2)/0.1)$. We also included various results from the greedy algorithm using different initial conditions in order to illustrate the non-convexity of the cost surface.

\begin{figure}
\begin{minipage}[b]{1.0\linewidth}
  \centering
  \centerline{\includegraphics[width=8.5cm]{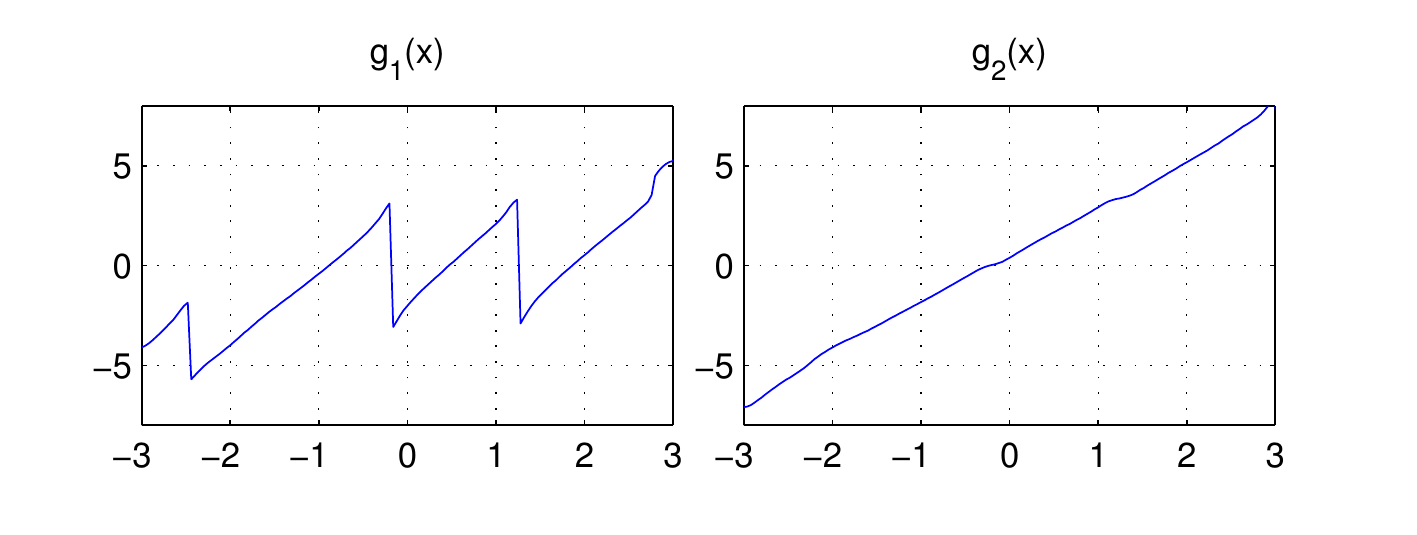}}
  \centerline{(a)}\medskip
\end{minipage}
\begin{minipage}[b]{1.0\linewidth}
  \centering
  \centerline{\includegraphics[width=8.5cm]{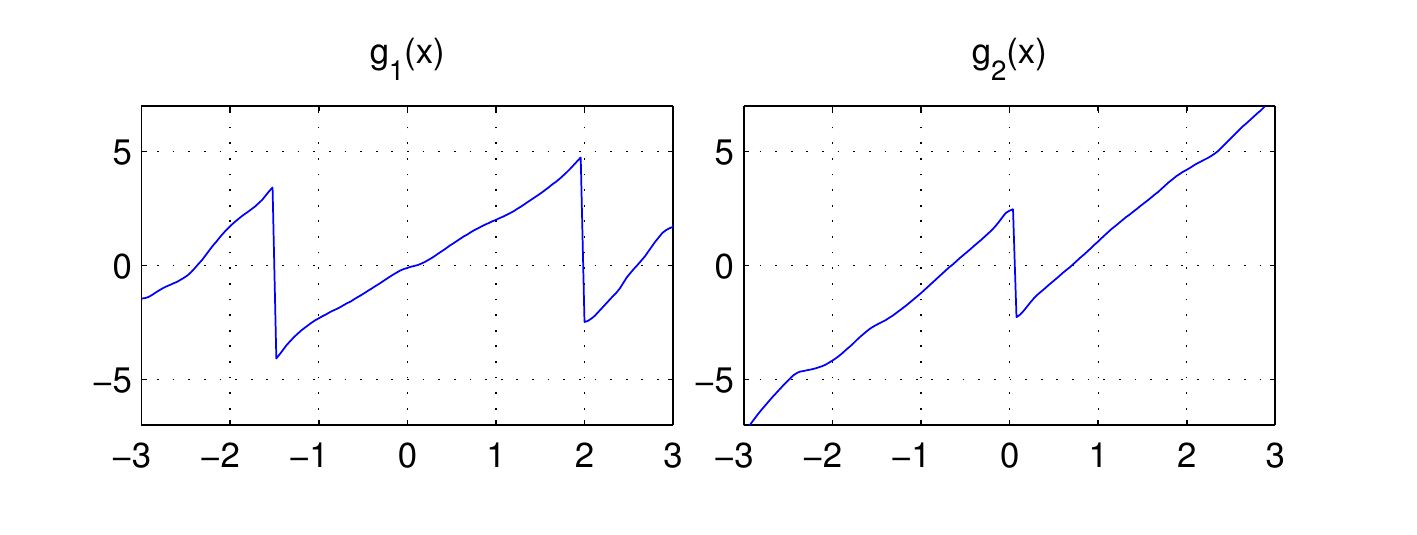}}
  \centerline{(b)}\medskip
\end{minipage}
\caption{Obtained encoders. (a) Individual power constraints, $P_1= 3.36$, $P_2=5.57$. (b) Total power constraint, $P_1=3.41$, $P_2=3.78$.}
\label{ex}
\end{figure}

Encoder mappings for two example settings are given in Figure \ref{ex}. In the first example we have individual power constraints. Note the similarity of this coding scheme to those found in prior work \cite{mehmetoglu_itw}, in the sense that the first encoder is a many-to-one mapping and the first source is recovered by using the output of the second channel. Intuitively, the second channel is used as side information since it is much more reliable due to higher power allocation. The second example in Figure \ref{ex} is obtained from a total power allocation setting. In this case, the powers of two encoders are close but not the same. Both encoders are many-to-one mappings in this case, that is, both channels are used as side information for each other in different source intervals. 

\section {Conclusions}
\label{conclusions}
In this paper we proposed an optimization method based on deterministic annealing ideas for optimizing analog distributed zero delay codes. Our method is  independent of initialization and provides results superior to the more ad hoc of noisy channel relaxation. The obtained mappings exhibit properties that are similar to digital Wyner-Ziv mappings. As part of future work, we seek to further investigate the theoretical properties of optimal mappings, as well  as application of the proposed ideas to related zero-delay analog coding problems such as multiple access channels.  

\bibliographystyle{IEEEtran}
\bibliography{ref}

\begin{thebibliography}{10}
\providecommand{\url}[1]{#1}
\csname url@samestyle\endcsname
\providecommand{\newblock}{\relax}
\providecommand{\bibinfo}[2]{#2}
\providecommand{\BIBentrySTDinterwordspacing}{\spaceskip=0pt\relax}
\providecommand{\BIBentryALTinterwordstretchfactor}{4}
\providecommand{\BIBentryALTinterwordspacing}{\spaceskip=\fontdimen2\font plus
\BIBentryALTinterwordstretchfactor\fontdimen3\font minus
  \fontdimen4\font\relax}
\providecommand{\BIBforeignlanguage}[2]{{%
\expandafter\ifx\csname l@#1\endcsname\relax
\typeout{** WARNING: IEEEtran.bst: No hyphenation pattern has been}%
\typeout{** loaded for the language `#1'. Using the pattern for}%
\typeout{** the default language instead.}%
\else
\language=\csname l@#1\endcsname
\fi
#2}}
\providecommand{\BIBdecl}{\relax}
\BIBdecl

\bibitem{goblick}
T.~Goblick~Jr, ``{Theoretical limitations on the transmission of data from
  analog sources},'' \emph{IEEE Transactions on Information Theory}, vol.~11,
  no.~4, pp. 558--567, 1965.

\bibitem{tocode}
M.~Gastpar, B.~Rimoldi, and M.~Vetterli, ``{To code, or not to code: {L}ossy
  source-channel communication revisited},'' \emph{IEEE Transactions on
  Information Theory,}, vol.~49, no.~5, pp. 1147--1158, 2003.

\bibitem{hekland}
F.~Hekland, G.~Oien, and T.~Ramstad, ``{Using 2: 1 Shannon mapping for joint
  source-channel coding},'' in \emph{Proceedings of the IEEE Data Compression
  Conference}, 2005, pp. 223--232.

\bibitem{karlsson2010optimized}
J.~Karlsson and M.~Skoglund, ``{Optimized low delay source channel relay
  mappings},'' \emph{IEEE Transactions on Communications}, vol.~58, no.~5, pp.
  1397--1404, 2010.

\bibitem{hu2011analog}
Y.~Hu, J.~Garcia-Frias, and M.~Lamarca, ``Analog joint source-channel coding
  using non-linear curves and mmse decoding,'' \emph{IEEE Transactions on
  Communications}, vol.~59, no.~11, pp. 3016--3026, 2011.

\bibitem{chen2011zero}
X.~Chen and E.~Tuncel, ``Zero-delay joint source-channel coding for the
  {G}aussian {W}yner-{Z}iv problem,'' in \emph{Proc. IEEE Int. Symp. on Inf.
  Theory}, 2011, pp. 2929--2933.

\bibitem{emrah_dcc10}
E.~Akyol, K.~Rose, and T.~Ramstad, ``{Optimized analog mappings for distributed
  source channel coding},'' in \emph{Proc. of IEEE Data Compression
  Conference}, 2010.

\bibitem{mehmetoglu_itw}
M.~S. Mehmetoglu, E.~Akyol, and K.~Rose, ``{A deterministic annealing approach
  to optimization of zero-delay source-channel codes},'' in \emph{Proc. of IEEE
  Inf. Theory Workshop}, 2013.

\bibitem{da}
K.~Rose, ``{Deterministic annealing for clustering, compression,
  classification, regression, and related optimization problems},''
  \emph{Proceedings of the IEEE}, vol.~86, no.~11, pp. 2210--2239, 1998.

\bibitem{gadkari1999robust}
S.~Gadkari and K.~Rose, ``{Robust vector quantizer design by noisy channel
  relaxation},'' \emph{IEEE Transactions on Communications}, vol.~47, no.~8,
  pp. 1113--1116, 1999.

\bibitem{Knagenhjelm}
P.~Knagenhjelm, ``{A recursive design method for robust vector quantization},''
  in \emph{Proc. Int. Conf. Signal Processing Applications and Technology},
  1992, pp. 948--954.

\end{thebibliography}
\end{document}